# Congestion and centrality in traffic flow on complex networks

Petter Holme
Department of Physics, Umeå University, 901 87 Umeå, Sweden

The central points of communication network flow has often been identified using graph theoretical centrality measures. In real networks, the state of traffic density arises from an interplay between the dynamics of the flow and the underlying network structure. In this work we investigate the relationship between centrality measures and the density of traffic for some simple particle hopping models on networks with emerging scale-free degree distributions. We also study how the speed of the dynamics are affected by the underlying network structure. Among other conclusions, we find that, even at low traffic densities, the dynamical measure of traffic density (the occupation ratio) has a non-trivial dependence on the static centrality (quantified by "betweenness centrality"), which non-central vertices getting a comparatively large portion of the traffic.

## I. INTRODUCTION

Networks are the underlying structure of many natural and man-made systems. In many cases the formation processes are very intricate, and the resulting networks can only be described as having both randomness and structure. Furthermore, dynamical systems can be confined to such complex networks—the flow of data packages in the Internet being one example. The interplay between the dynamical systems and their underlying network structure is vast and intriguing area of research. The main topic of this paper is the relation between centrality assessed from the static network structure measured in simulations of some simple traffic flow models. We also study how the speed of the traffic flow is affected by the network structure (by tuning model parameters).

The concept of centrality in networks can be traced back to the 19th century (1). Following the work of Freeman (2) many centrality measures, designed to capture different aspects of the centrality concept, has been proposed within study of social networks. A particular favorite in recent network literature is the betweenness centrality (3). Roughly speaking the betweenness of a vertex $v$ is the number of shortest paths between all pairs of vertices that passes $v$. Routing communication along the shortest paths is of course beneficial for speed and economy, but if there is a limit to the vertex load and network traffic is heavy, congestion is a threat to central vertices. In real network flow the overall efficiency can be improved if some pairs settles with sub-optimal routes (so called load balancing). This is obtained implicitly in Internet routing protocols (such as the Border Gateway Protocol (4)) by using timings from earlier routings. Still, vertices of high betweenness can be assumed to convey a large amount of traffic; and the more dilute the traffic is, the better the agreement. In fact, several papers has used betweenness as a measure for the vertex load in dynamical communications systems (5–9). In this paper we investigate the relationship between the betweenness centrality and congestion in simple particle hopping models for traffic flow. The underlying networks are constructed by an extension (10) of the Barabási-Albert (BA) algorithm (11). This construction enables us to study networks of a similar skewed degree distribution as communication networks such as the Internet (12–14), and also to monitor the effects of clustering—the density of triangles of the network—which is known to be high in e.g. the large scale of the Internet (15).[1]

## II. DEFINITIONS

### A. Graph theoretical preliminaries

We model the networks as graphs $G = (V, E)$ where $V$ is the set of vertices and $E$ the set of edges (unordered pairs of vertices). A path is a sequence of vertices, such that there is an edge in $E$ between all consecutive pairs of vertices. A geodesic between $u, v \in V$ is a path containing fewest possible number of vertices; the number of edges of a geodesic between $u, v \in V$ is called distance $d(u, v)$.

### B. Betweenness centrality

As mentioned the betweenness centrality of a vertex is a count of the number of geodesics passing that vertex. More precisely, let $v$ be a vertex in $V$, then the betweenness centrality $C_B(v)$ is defined as

$$C_B(v) = \sum_{u\in V} \sum_{w\in V\setminus\{u\}} \frac{\sigma_{uw}(v)}{\sigma_{uw}} , \tag{1}$$

where $\sigma_{uw}$ is the number of geodesics (shortest paths) between $u$ and $w$, and $\sigma_{uw}(v)$ is the number of geodesics between $u$ and $w$ that passes $v$.

[1] It should be noted that there are yet other structures e.g. skewed degree-degree correlations (16; 17) that may affect the network flow. This is an interesting question for further studies but is not addressed in the present paper.

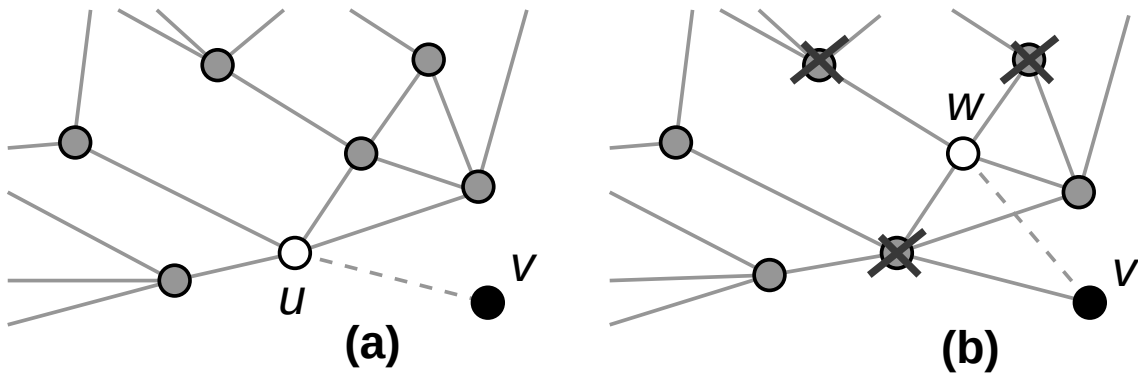

FIG. 1 The construction of the clustered scale-free network in Sec. II.C. (a) In the preferential attachment step for the newly added vertex $v$ (denoted as filled black circle), the white vertex $u$ is chosen with the probability proportional to its degree (the dotted line represents the new edge). (b) In the triangle formation step an additional edge (dotted line) is added to a randomly selected vertex $w$ in the neighborhood $\Gamma_u$ of the vertex $u$ chosen in the previous preferential attachment step in (a). The crossed out vertices are not allowed since they are not in $\Gamma_u$. Without the triangle formation step (i.e. if $m_t = 0$), the clustered scale-free model reduces to the original BA model of scale-free networks.

### C. Networks with power-law degree distribution and tunable clustering

To be able to monitor the traffic density as a function of betweenness we need networks with a broad betweenness distribution. This can be obtained by using scale-free networks as generated by the Barabási-Albert (BA) model (11).[2]

To generate networks with both an emerging scale-free degree distribution and tunable clustering one can use the following extension of the BA model (11): Starting from $m_0$ vertices and no vertices one grows the network iteratively by adding one vertex $v$ and $m$ edges adjacent to $v$ at each timestep. The attachment of the ends opposite to $v$ to a vertex $u$ of the old part of the network is done with a probability in proportion to the degree of that vertex, this so called preferential attachment gives rise to multiplicative effects that induces an emergent scale-free degree distribution (18; 19).

There are more types of structure in real world communication networks such as the Internet—another example is clustering, or high density of triangles. Let $p(n)$ be the number of paths of length $n$ (a path is a $n$-tuple of vertices $(v_1, v_2, \cdots, v_n)$ such that $v_i$ and $v_{i+1}$ are adjacent for any $i \in [1, n]$), and let $c(n)$ be the number of circuits of length $n$ (i.e. $n$-tuples such as mentioned above where also $v_n$ is connected to $v_1$). Then the clustering can be quantified as

$$C = \frac{c(3)}{p(3)} \qquad (2)$$

where $C$ is called clustering coefficient (20). $C$ lies strictly in the interval $[0, 1]$; where $C = 0$ means that the graph contains no triangles, and $C = 1$ if every path of length three is a triangle.

The Barabási-Albert model is known to have vanishing clustering in the large $N$ limit. To introduce high clustering we follow Ref. (10) and do not add all edges by preferential attachment, but with a probability $p_t = m_t/(m-1)$ replaces a preferential attachment step by a "triangle formation" step: If a new vertex $v$ has attached to a vertex $u_1$ by preferential attachment in the preceding step, one selects one of $u_1$'s neighbor uniformly at random to get another vertex $u_2$ to attach to. The first edge of a new vertex is, however, always added with preferential attachment. See the illustration in Fig. 1.

### D. Particle hopping dynamics

Particle hopping is a widely used modeling technique in vehicular flow theory[3] In this approach a road is represented as a string of cells. These cells can be empty or occupied by exactly one particle. Modeling traffic flow by such a cellular automaton formalism[4] is believed to give minimal models for the universal behaviors of traffic flow. We note that cellular automata formalism has been used to model packet flow in the Internet (26; 27). We aim for yet simpler models that still allow congestion and propose three dynamical rules for updating the particles, all based on the particle hopping models mentioned above—a fraction $f$ of the vertices occupied by particles, a one-particle limit to each vertex, and the following dynamics: The $Nf$ particles are updated sequentially—so called Little dynamics. To be able to compare different particle densities we use the number of updates divided with the number of particle as a time unit, i.e. one unit of time is the average number of iterations between the updates of one particle.

Each particle has a defined start and stop position, and is thus the full situation of the particle is represented by a triple $p_i(t) = (v(t), v_{\text{start}}, v_{\text{target}})$ where $v$ is the present position, $v_{\text{start}}$ is the starting position and $v_{\text{target}}$ is the target site. When a target site is reached, a new target site is randomly chosen and the old target vertex becomes the new $v_{\text{start}}$. The different dynamical updating rules for particle moves we consider are:

**Random walk (RW).** In this case the particle choose an unoccupied neighbor site uniformly at random. If no unoccupied neighbors exists the particle rests.[5]

---

[2] There is no known way to sample all graphs of a given degree distribution. The advantage with the BA model is that it is the mostly studied and lacks structural-biases such as non-zero degree-degree correlations.

[3] This approach was pioneered by D. L. Gerlough in the 1950s (21). For reviews, see Refs. (22–25).

[4] Strictly speaking one also have to require the dynamics to be updated in parallel for a particle hopping model to qualify as a cellular automaton.

[5] The relation between regular random walk dynamics and the un-

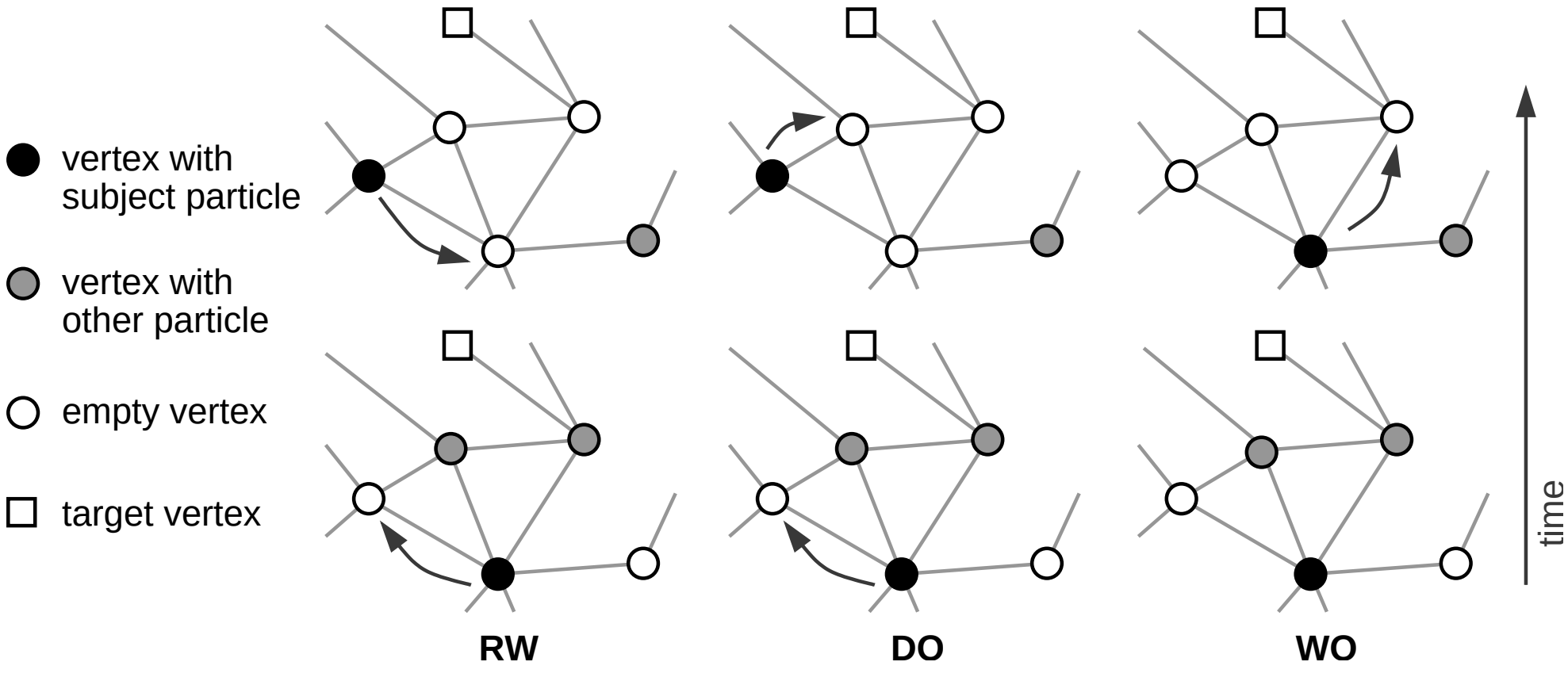

FIG. 2 Illustration of the three different update rules. In the first move the RW dynamics chooses any free vertex, the DO dynamics moves away from the target, and WO waits one time step. In the next step both DO and WO can continue towards the target.

**Detour at obstacle (DO).** The particles chooses randomly among the neighbors closest to the target; it do jump even if the distance stays constant or increases. This dynamics is meant to be a simple dynamics that works quite effectively both in sparsely and densely trafficked networks. With no particles meeting the routes are simply the geodesics. If a traffic jam occurs the particles do their best to get out of it; it might of course be better to wait, but there is no way of telling this, for a particle, without knowing the intended routes of the other particles.

**Wait at obstacle (WO).** Like the DO case the particle starts by trying to get closer to the target. If no neighbor vertex closer to the target exists the particle rests. In this dynamics the particles necessarily travels along geodesics. There is of course a possibility that particles heading in opposite directions ends up dead-locked. In this sense this dynamics is unrealistic as a model for real communication systems. On the other hand, in a dilute system the avoidance of backing particles could, on average, speed up the dynamics. Furthermore the WO case serves as the most extreme case of path-length economy (and is it that sense the one closest to the idea of betweenness centrality).

These updating rules are illustrated in Fig. 2.

For one run of the simulation we generate a networks and a starting configuration for the set of particles. A run has ended when either all particles has reached their target or all particles has remained fixed in their last update step. Data for finding time are gathered for the particles time to reach their initial target configurations. Data for the other principal dynamical quantity we measure—the occupation ratio $w$ defined as the fraction of time steps that a vertex is occupied by a particle—are gathered throughout a whole run. In each run we use new random seeds both for generating networks and the particle dynamics. All figures are averaged over 1000 runs.

## III. SPEED OF THE DYNAMICS

This section gives an overview over how the speed of the dynamics, as measured by the average finding time $t_f$, is affected by the network's structure and size. Some preliminary conclusions about the interplay between the particle dynamics and the underlying network structure will also be drawn.

Fig. 3(a) shows the average finding time for pairs $u, v \in V$ at a distance $d(u, v) = 4$ as a function of particle density $f$ for different clustering and dynamical updating rules. (The use of $d(u, v) = 4$ and one system size $N = 200$ will be discussed below.) For all particle densities the WO dynamics are fastest and is almost $f$-independent. This is not the whole truth however as the WO dynamics can end up in a dead-locked state. The ratio of runs when this happens $\phi$, displayed in Fig. 3(b), rapidly decreases as $f$ tends to unity. Evidently the WO strategy would be useless in a real system (with the one-particle-a-vertex restriction). The interesting information is rather that it is most often true that a particle reaches the target if and only if there is no dead-lock, i.e. a slowly moving state of the WO system is very rare. In the dilute limit ($f \to 0$) the DO and WO finding times converge to the distance between the start and stop vertices (remember that the particles not interacting with others travels along geodesic with these dynamics). When $f = 1$ all particles are trivially immobile; hence as $f \to 1$ the finding time for both the RW and DO dynamics diverge (see Fig 3(a)) and the WO success ratio goes to zero. RW remains the slowest dynamics throughout the $f$-axis, but as $f \to 1$ the DO and RW finding times converges. This can be interpreted as when the particle density increases, the particle increasingly often has to settle for a move

---

derlying network structure has been studied in Refs. (28; 29).

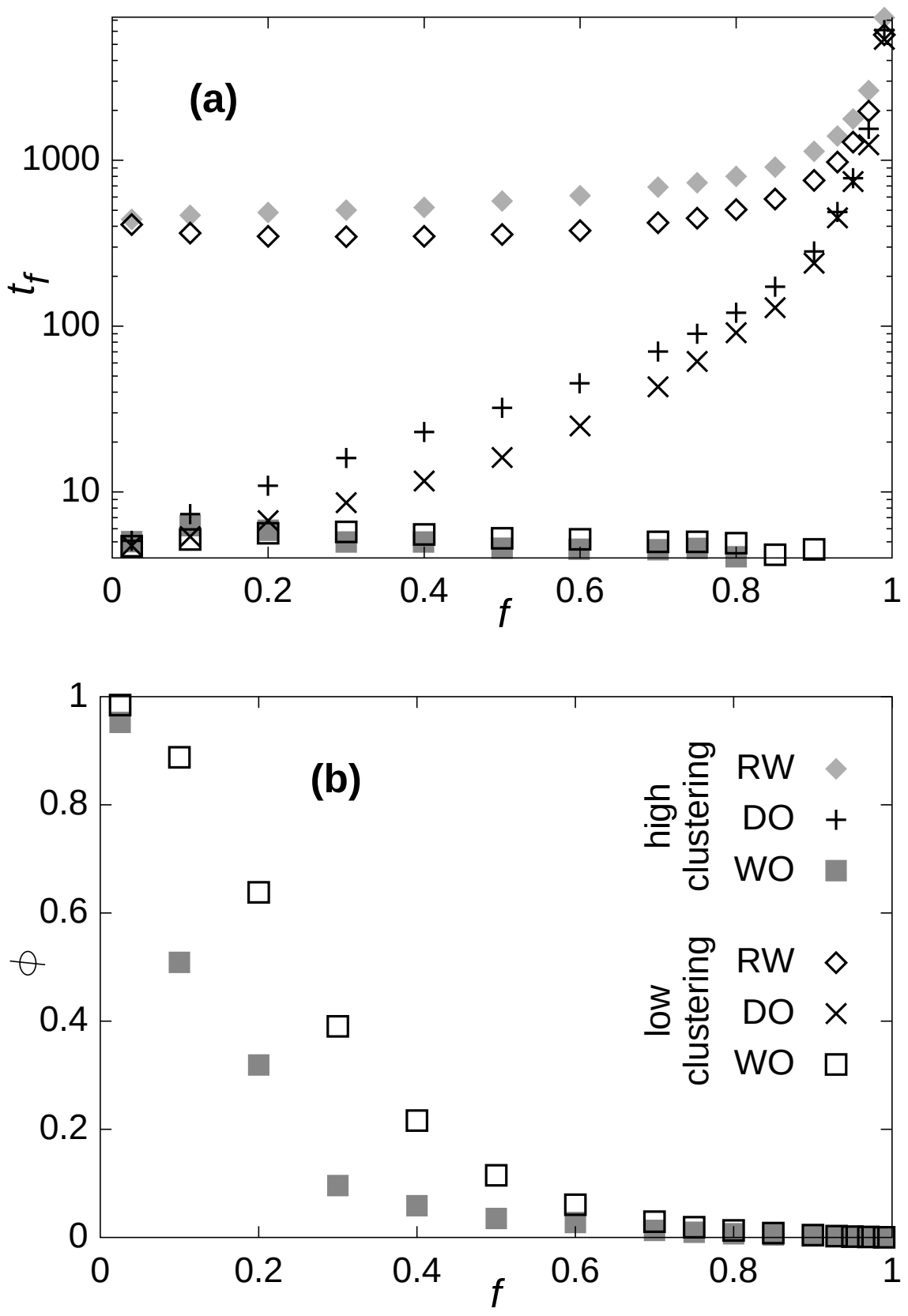

FIG. 3 The traffic-density dependence of speed of the dynamics: Average finding $t_f$ time for pairs at distance four as a function of particle density for the three different types of dynamics, and both low ($m_t = 0$ giving $C \approx 0.056$ for the present values of $m$, $m_0$ and $N$) and high clustering ($m_t = 2.8$ giving $C \approx 0.24$). (b) shows the fraction of pairs that reaches their targets $\phi$ as a function of $f$ in the WO dynamics. For the RW and DO updating rules $\phi$ is strictly unity. The other model parameters are $N = 200$ and $m = m_0 = 3$. Error bars are displayed only if larger than system size.

that does not change, or increases, the distance to the target; thus, becoming increasingly similar to the RW strategy. In the extreme case with $N - 1$ particles, a particle will move to the free vertex if it is in the free vertex' neighborhood, otherwise it will rest, the same as for the RW case.

Next we investigate the finite size scaling behavior of the system. From Fig. 4 we observe that there are no emerging discontinuities as $N \to \infty$. In fact, the RW and DO curves are all scaling algebraically towards infinity as $N$ grows, and similarly the success ratio in the WO case is going to zero as a power-law (see Fig. 5). From a statistical mechanics point of view, i.e. taking $N \to \infty$, this means that the systems we consider always are in a congested state and does not undergo any phase transitions for the $f$ range in question[6]. Since our main topic is the relation between flow density and betweenness centrality in finite life-like communication networks out of the trivial free-flow state this is good news—we can draw qualitatively correct conclusions from moderate system sizes, which justifies the conclusions to come (and above). That a network such as the Internet experiences spatially and temporally limited congestion is a well known phenomena (30–32).

We briefly mention the finding time as a function of distance (displayed in Fig. 6). These curves are bounded from below by the target distance. In the low density limit, the DO and WO dynamics lies expectedly very close to this free-flow bound. Even if all curves lies above the $t = d$ limit, the curves seems to bend down slightly for increasing $d$. This is presumably due to when a pair is separated by a distance close to the diameter (maximum pair distance) the shortest paths passes relatively many non-central vertices[7], which are less often occupied, and hence gives a (quite counter-intuitive) bias towards short finding times for large separations. We hope future studies will sort out the mechanism in detail. The lack of emerging singularities in Fig. 6 justifies the conclusions drawn from Fig. 3.

Finally we turn to the effect of high clustering. For both the RW and DO dynamics the finding time increases with clustering for the whole range of particle densities (see Figs. 3 and 6). This slowing of the dynamics induced by clustering can be explained from the wider betweenness distribution as found in Ref. (34). If the occupation ratio is an increasing function of betweenness (which is, as we will see in the next section, indeed the case), then the traffic will easily be congested at the vertices of highest betweenness (of which there are more in the clustered networks) and thus the overall dynamics will be slower.

## IV. RELATION BETWEEN CONGESTION AND BETWEENNESS CENTRALITY

In this section we turn to the relationship between a vertex' average occupation ratio and betweenness centrality. The results are strengthened by finite size scaling and the correlation coefficient between $w(v)$ and the maximal $w$ in the neighborhood of $v$.

Fig. 7 shows the occupation ratio as a function of betweenness centrality for the three different dynamics in high and low particle densities, and high and low clustering. In the $f \to 0$ limit the particles travels along geodesics in the DO and WO dynamics. In this case, assuming no collisions, $w$ is proportional to $f C_B$. But already at $f = 0.1$ (see Fig. 7(a) and (b)) $w$'s linear $C_B$-dependence is broken and the $w(C_B)$ curve turns convex ($\partial^2 w/\partial C_B^2 < 0$). This is a clear sign of traffic-jams at the network's more highly connected vertices. For small $C_B$ the occupation ratio reaches a plateau, as mostly easily seen for high particle densities (as in Figs. 7(c) and (d)). This can be explained as a direct result of the zero degree-degree correlations in BA model network (16): Assuming that traffic-jams preferably are centered around vertices of high degree (that most likely have high betweenness (34)), neighbors of high-degree vertices will also have a high occupation number; but since the degrees of neighboring vertices are uncorrelated this will give a constant contribution to the $w(C_B)$-curves ($C_B$ is almost proportional to degree for the model networks in question (34)). This effect is hard to anticipate from pure graph theoretical considerations. Considering that only a minor fraction of the vertices belongs to the high $C_B$-tail (with a higher $w$) the betweenness distribution does not give the complete picture, even for a low particle density. The small qualitative difference between low (Fig. 7 (a) and (c)) and high (Fig. 7 (b) and (d)) clustering is that vertices with high $C_B$ have larger $w$ values. Thus, clustering does not only give more vertices of high betweenness, points with high betweenness have larger occupation ratios.

[6] If $f = 0$, the system is not in a congested state. The phase transition takes place for $f = 0$ or $f$ very close to 0.

[7] Since the most central vertices acts as hubs that gives the graph its small-world characteristics (i.e. a slower than polynomial increase of the average geodesic length) (33).

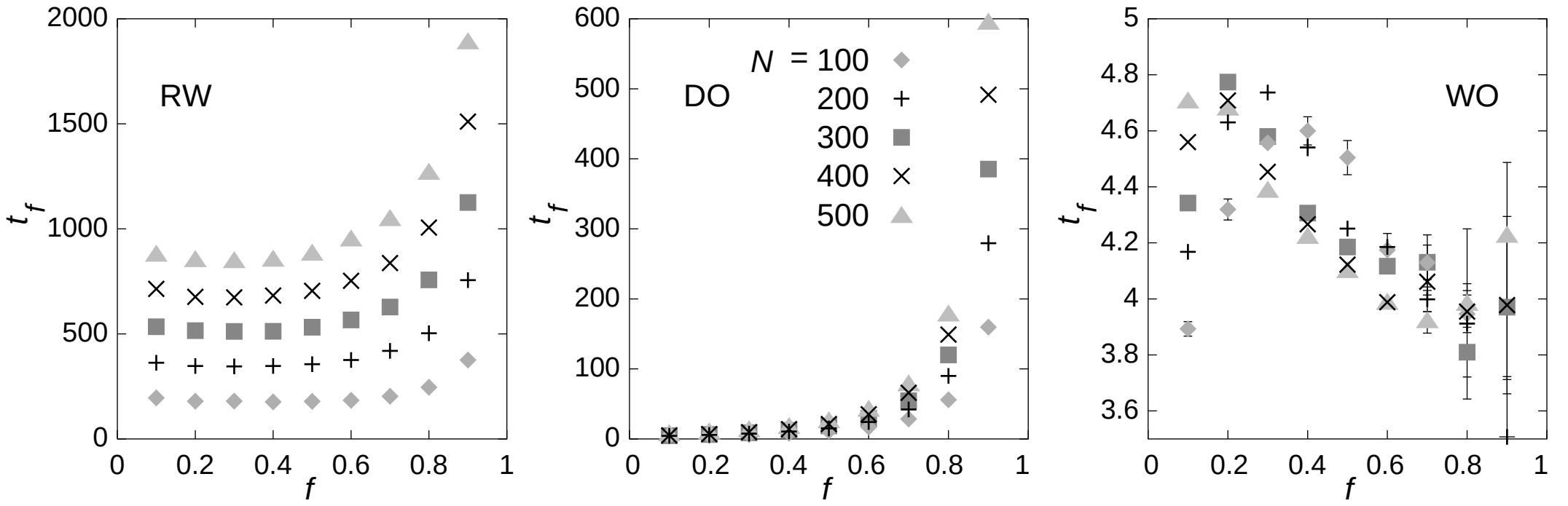

FIG. 4 Finding time of vertex-pairs at distance four for different system sizes. The system parameters are $m_t = 0$, $m = m_0 = 3$. The lack of emerging singularities for $0.1 \leq f \leq 0.9$ implies that qualitative conclusions from moderate system sizes will hold (in this region of $f$) for arbitrary finite system sizes.

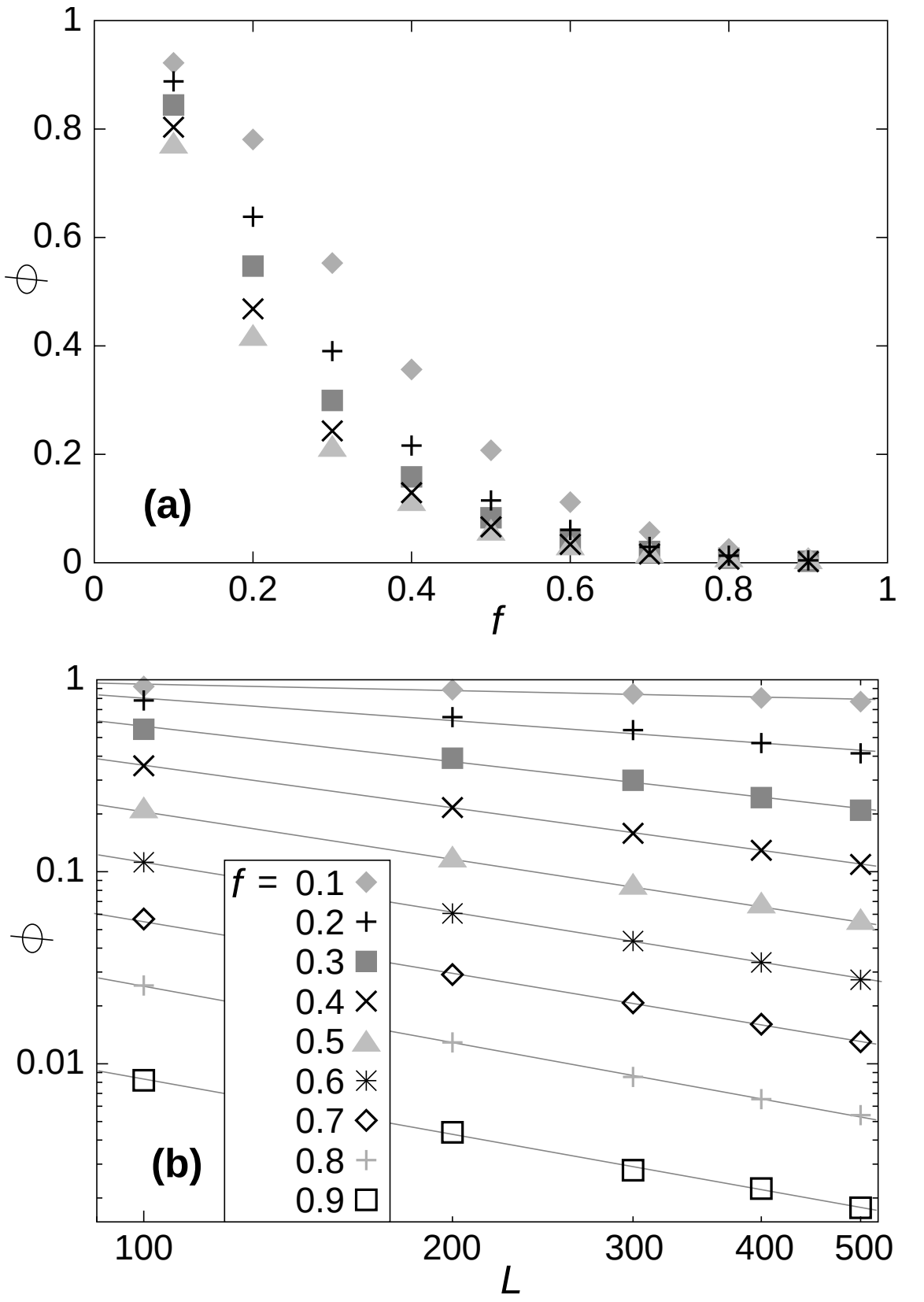

FIG. 5 Finite size scaling of the WO dynamic's finding probability $\phi$. (a) shows $\phi$ as a function of $f$ while (b) displays $\phi$ as a function of system size. The lines are curve-fits to a power-law form. The symbols in (a) are the same as in Fig. 4. Just as the RW and DO dynamics finding times goes to infinity (as seen in Fig. 4), $\phi$ goes to zero, which shows that all three dynamics are (from a statistical mechanics point of view) in a congested state.

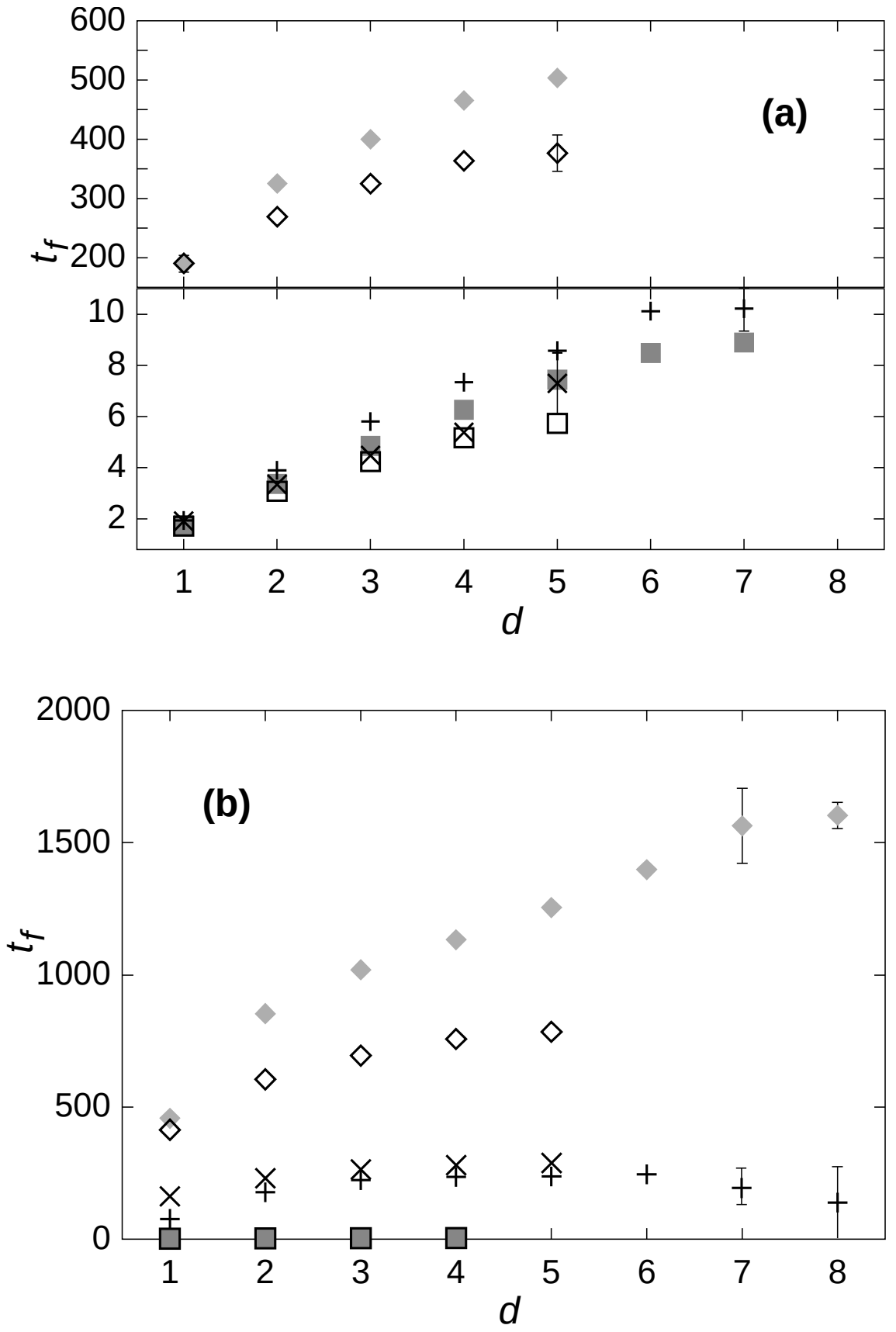

FIG. 6 Finding time as a function of geodesic distance to the target. (a) shows the situation for low particle density $f = 0.1$. (b) shows the corresponding situation for high particle density $f = 0.9$. Symbols are the same as in Fig. 3.

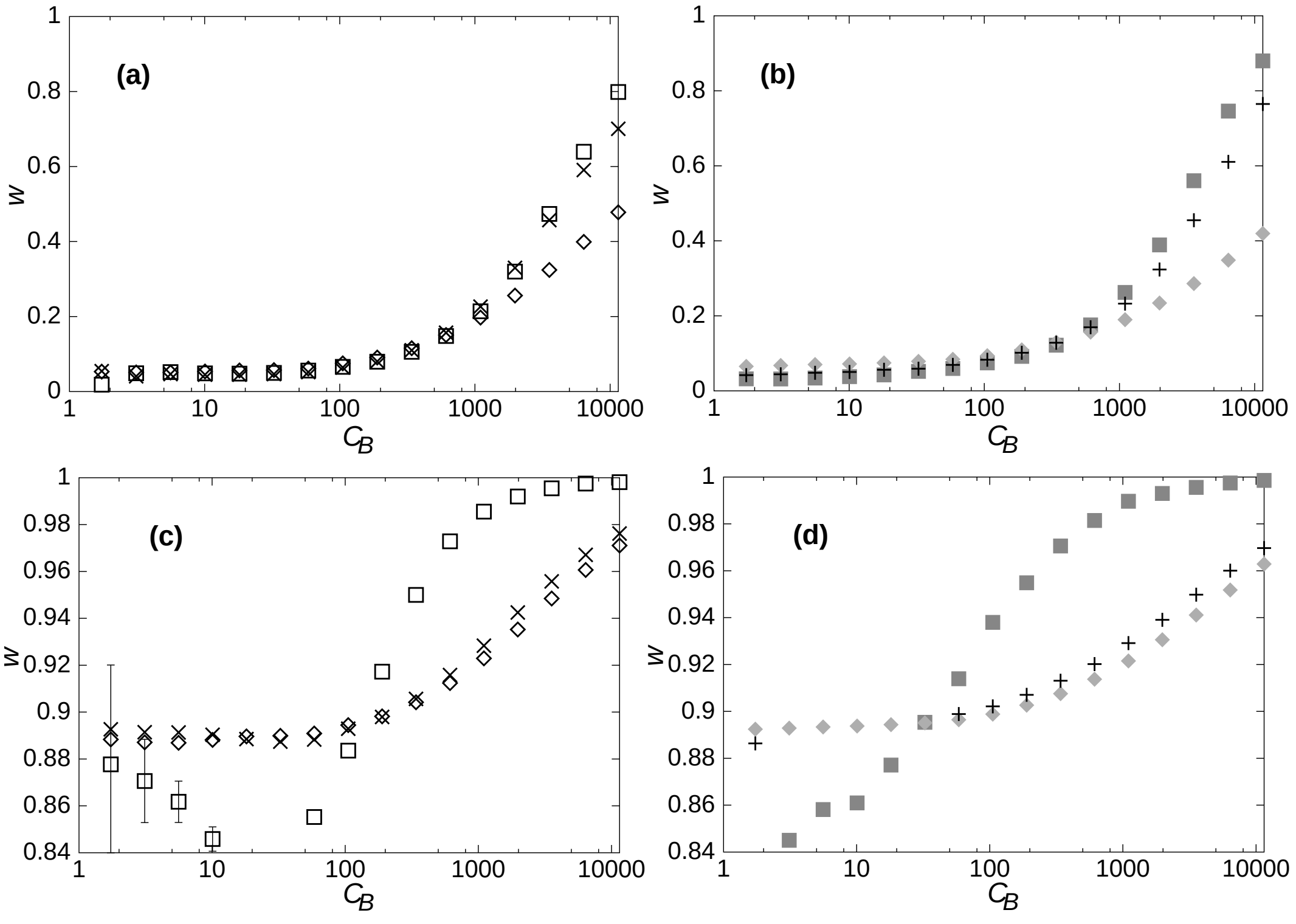

FIG. 7 Occupation time as a function of betweenness. (a) and (b) shows the three dynamics at the low particle density $f = 0.1$; (c) and (d) shows the corresponding situation for a high particle density $f = 0.9$. (a) and (c) shows networks of low clustering $m_t = 0$ (giving $C \approx 0.056$) while (b) and (d) have high clustering $mt = 2.8$ (giving $C \approx 0.24$). The other model parameters are $N = 200$ and $m = m_0 = 3$. Symbols are the same as in Fig. 3.

Another immediate observation is that, for small $f$, the fastest dynamics (DO and WO) have a higher occupation ratio for the most central vertices compared to the slower RW updating strategy. For large particle densities the WO dynamics seldom terminates so DO and WO are the only usable strategies, and these have a fairly high occupation number for non-central vertices. The explanation of this can be done by an allegory from the common urban traffic experience—during the off-peak hours the fastest routes are the shortest and goes through the largest streets, whereas during the rush hours an experienced driver can save minutes by navigating through back-alleys. In the dense limit, particles (as well as cars) are forced to make a detour around the traffic-jams.

Since the finding time behaves smoothly throughout our parameter ranges we expect a moderate system size, like in Fig. 7 to give a qualitatively correct behavior. This is confirmed in Fig. 8 where $w$ is plotted against the normalized betweenness $c_B(v) = C_B(v)/\langle C_B\rangle$. The curves converge pointwise and is, for the studied sizes, almost overlapping (as expected from the small size dependence seen in Fig. 3).

In a congested traffic situation one can expect that a vertex is likely to be jammed if a neighbor is jammed. To test this we measure the $w$-$W$ correlation coefficient, where $W$ is the maximum occupation ratio of $w$'s neighbors ($W(v) = \max_{u\in\Gamma_v} w(u)$):

$$r_{wW} = \frac{\langle w(v)\, W(v)\rangle - \langle w(v)\rangle\langle W(v)\rangle}{\sqrt{\langle w(v)^2\rangle - \langle w(v)\rangle^2}\,\sqrt{\langle W(v)^2\rangle - \langle W(v)\rangle^2}} \qquad (3)$$

where the average $\langle\cdot\rangle$ is over all vertices $v \in V$. $r_{wW}$ is the covariance normalized to lie strictly in the interval $[-1, 1]$. A positive value means that there is a positive correlation between $w$ and $W$; i.e. if much traffic passes $v$ then it is likely to have a neighbor with high occupation ratio (and vice versa). A large positive $r_{wW}$ would then support the idea that there are congestion centers in the network; and since a large $w$ implies a large $C_B$ (more or less) it would support the intuitive idea that traffic jams are centered around vertices of high betweenness.

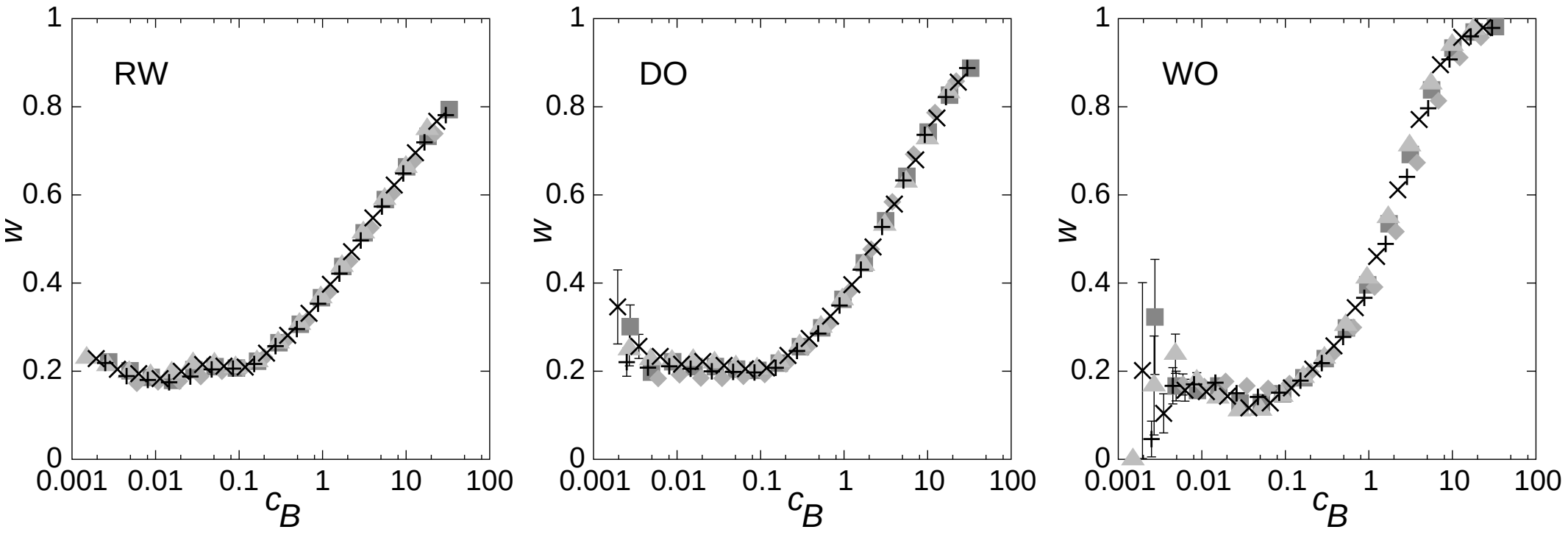

FIG. 8 Occupation time as a function of normalized betweenness for different $N$. The other model parameters are $f = 0.3$, $m_t = 0$, and $m = m_0 = 3$. The symbols are the same as in Fig. 4. The curves overlap as $N$ grows indicating that the qualitative picture is the same for arbitrary large sizes.

As seen in Fig. 9, it is indeed the case that all dynamics shows more or less positive $r_{wW}$ values. Fig. 9(a) shows that $r_{wW}$ is fairly $N$-independent. Fig. 9(b) gives the asymptotic values as $N \to \infty$. Just as expected (from the discussion in context of Fig. 3) RW and DO seems to converge in the dense limit and WO $\approx$ DO in the sparse limit. WO gives the lowest correlation coefficient; this can be explained by the dead-locks appearing in this dynamics—local dead-locked configurations (of a few particles) can appear even at less central vertices; such a situation appearing quite long before the whole system is dead-locked gives a negative contribution to $r_{wW}$. The RW dynamics has a rather high value for small $f$ which is can be explained as follows: For a one-particle system, the expectation value for $w$ is higher for vertices close to $v_{\text{start}}$; so, after an average run, the region close to $v_{\text{start}}$ will have high values of both $w$ and $W$ whereas the remote part of $G$ has a low traffic intensity and thus low $w$ and $W$. High densities is almost a dual to low densities for the RW updating rule—with $N-1$ particles the vacancy follows the same dynamics as a solitary particle, the only difference is that one run lasts much longer (but the argument for a high $r_{wW}$ is the same). More than one particle weakens the discussed effect giving a concave shape of the $r_{wW}$ curve.

The only parameter not discussed so far is the average degree in the graph, as controlled by $m$. As the density of edges increases the there are more direct routes releasing some pressure off the most connected vertices. As a result, the betweenness distribution becomes more narrow and effects of congestion will become smaller, but there will presumably not occur any qualitative changes. Preliminary simulation results strengthens this conjecture.

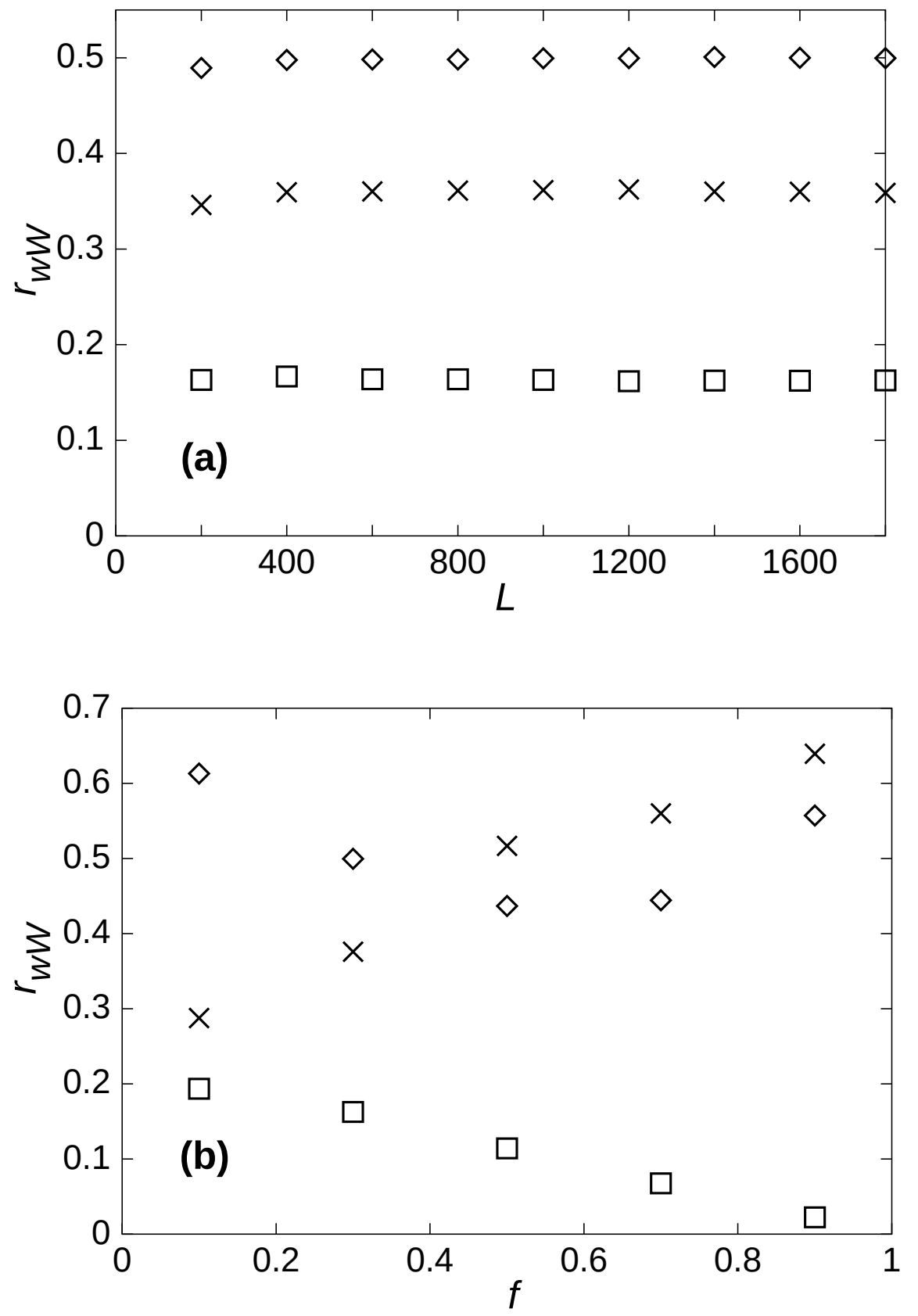

FIG. 9 (a) The correlation coefficient $r_{wW}$ as a function of system size for $f = 0.3$. (b) The asymptotic value of $r_{wW}$ as $N \to \infty$ as a function of $f$. The other parameter values are $m = m_0 = 3$, and $m_t = 0$. The symbols are the same as in Fig. 3. The positive values shows that the probability for a vertex to be congested increases if a vertex in its neighborhood is congested.

## V. SUMMARY AND CONCLUSIONS

We have used particle hopping models adapted from vehicular traffic flow theory as minimal models for a communication systems that can be congested. In these models a number of particles move along the edges of the graph with the restriction that not more than one particle at a time can occupy a vertex. Three different dynamics for updating particle positions were considered: RW (random walk), where particles move at random. DO (detour at obstacle), where particles move to the neighboring vertex closest to its target, and moves even if the position gets worse. WO (wait at obstacle), that works like DO, but where particles rests if the position would get worse. To make the investigation of the influence of the underlying network structure on these dynamical systems, we need networks with (just as many real-world communication networks) a wide spread in the betweenness distribution. This is obtained by using the BA model for scale-free networks. To test the effect of clustering we also use an extended version of the BA model.

For completed connections (from start vertex to target vertex) the speed of the updating rules ranks as RW < DO < WO, for all parameter values, with RW ≈ DO for high particle densities and DO ≈ WO for sparse traffic. However the WO dynamics ends up in a dead-lock with probability that increases with the particle density. High clustering slows down the dynamics, but does not lead to any qualitative change. The slowing down can be explained since higher clustering stretches the tail of the betweenness distribution (34), and high betweenness is likely to form congestion centers that slows down the dynamics.

Even at very low particle concentrations, the picture from betweenness distribution changes. In particular vertices over a wide range of small and medium betweennesses all have quite constant occupation ratios. In the context of communication systems design this means that the capacity of a vertex in the network cannot be estimated by its own betweenness alone—having a neighbor prone to congestion is probably worse (give a higher occupation ratio) than having a medium high betweenness. This statement is explicitly illustrated by a strong positive correlation between the occupation ratio $w$ of a vertex $v$ and the maximum occupation ratio $W$ in $w$'s neighborhood.

In a more general perspective we believe that this work illustrates one of the key-areas of future complex systems studies—the interplay between the behavior of individual entities and the structure in which they interact. Even if one can imagine more sophisticated measures of the static structure than we use in the present work[8], synergetic effects can be considerable and even counter-intuitive (cf. Ref. (36)).

## Acknowledgements

The author thanks Beom Jun Kim for constructive comments. This work was partially supported by the Swedish Natural Research Council through Contract No. F 5102-659/2001.

---

[8] In the case of vertex load in a communication network with load balancing the "flow betweenness" (35) is probably more relevant than betweenness. The flow betweenness takes all paths, not only the shortest, into account; but suffers from being computationally quite intractable.

## References

[1] Jordan, C., *J. Reine Angew. Math.* **70**, 185 (1869).
[2] Freeman, L. C., *Soc. Networks* **1**, 215 (1979).
[3] Freeman, L. C., *Sociometry* **40**, 35 (1977).
[4] Huitema, C., *Routing in the Internet* 2nd ed. (Prentice Hall, Upper Saddle River NJ, 2000).
[5] Goh, K.-I., Kahng, B., and Kim, D., *Phys. Rev. Lett.* **87**, 278201 (2001).
[6] Holme, P. and Kim, B. J., *Phys. Rev.* **E65**, 066109 (2002).
[7] Szabó, G., Alava, M., and Kertész, J., *Phys. Rev.* **E66**, 026101 (2002).
[8] Goh, K.-I., Oh, E. S., Jeong, H., Kahng, B., and Kim, D., *Proc. Natl. Acad. Sci. USA* **99**, 12583 (2002).
[9] Holme, P., *Phys. Rev.* **E66**, 036119 (2002).
[10] Holme, P. and Kim, B. J., *Phys. Rev.* **E65**, 026107 (2002).
[11] Barabási, A.-L., and Albert, R., *Science* **286**, 509 (1999).
[12] Faloutsos, M., Faloutsos, P., and Faloutsos, C., *Comput. Commun. Rev.* **29**, 251 (1999).
[13] Kumar, R., Rajalopagan, P., Divakumar, C., Tomkins, A., and Upfal, E., in *Proceedings of the 19th Symposium on Principles of Database Systems* (Association for Computing Machinery, New York, 1999).
[14] Broder, A., Kumar, R., Maghoul, F., Raghavan, P., Rajagopalan, S., Stata, R., Tomkins, A., and Wiener, J., *Comp. Networks* **33**, 309 (2000).
[15] Vazquez, A., Pastor-Satorras, R., Vespignani, A., *Phys. Rev.* **E65**, 066130 (2002).
[16] Newman, M. E. J., *Phys. Rev. Lett.* **89**, 208701 (2002).
[17] Maslov, S., Sneppen, K. and Zaliznyak, A., "Pattern Detection in Complex Networks: Correlation Profile of the Internet", cond-mat/0205379, May 2002.
[18] Simon, H. A., *Biometrika* **42**, 425 (1955).
[19] de S. Price, D. J. *J. Amer. Soc. Inform. Sci.* **27**, 292 (1976).
[20] Newman, M. E. J., Strogatz, S. H., and Watts, D. J., *Phys. Rev.* **E64**, 026118 (2001).
[21] Gerlough, D. L. in *Proceedings of the 35th annual meeting* (Highway Research Board, Washington DC, 1956).
[22] Nagel, K., *Particle Hopping Models and Traffic Flow Theory*, technical report no. LA-UR 95-2908 (Los Alamos National Laboratory, 1995).
[23] Chowdhury, D., Santen, L., and Schadschneider, A., *Phys. Rep.* **329**, 199 (2000).
[24] Schweitzer, F., *Brownian Agents and Active Particles* (Springer, Berlin, 2001).
[25] Helbing, D., *Rev. Mod. Phys.* **73**, 1067 (2001).
[26] Huisinga, T., Barlovic, R., Knospe, W., Schadschneider, A., Schreckenberg, M., *Physica* **A294**, 249 (2001).
[27] Liu, F., Ren, Y., and Shan, X. M., *Acta Phys. Sinica* **51**, 1175 (2002).
[28] Lahtinen, J., Kertész, J., and Kaski, K., *Phys. Rev.* **E64**, 057105 (2001).
[29] Tadić, B., *Eur. Phys. J.* **B23**, 221 (2001).
[30] Takayasu, M., Takayasu, H., and Sato, T., *Physica* **A233**, 824 (1996).
[31] Fukuda, K., Takayasu, H., and Takayasu, M., *Advances in Performance Analysis* **2**, 21 (1999).
[32] Fukuda, K., Takayasu, H., and Takayasu, M., *Fractals* **7** 23 (1999).
[33] Amaral, L. A. N., Scala, A., Barthélémy, M., Stanley, H. E., *Proc. Natl. Acad. Sci. USA* **97**, 11149 (2002).
[34] Holme, P., Kim, B. J., Yoon, C. N., and Han, S. K., *Phys. Rev.* **E65**, 056109 (2002).

[35] Freeman, L. C., Borgatti, S. P., and White, D. R., *Soc. Networks* **13**, 141 (1991).
[36] Kim, B. J., Trusina, A., Holme, P., Minnhagen, P., Chung, J. S., and Choi, M. Y., *Phys. Rev.* **E66**, 021907 (2002).